\documentclass[submission]{eptcs}
\providecommand{\event}{DORIC-MM 2021} 
\usepackage{graphicx}

\usepackage[sorting=none]{biblatex}

\bibliography{mambo_reference}

\title{Introducing MAMBO: Materials And Molecules Basic Ontology}
\author{Fabio Le Piane
\institute{Istituto per lo Studio dei Materiali Nanostrutturati (ISMN)}
\institute{Consiglio Nazionale delle Ricerche (CNR)\\
Bologna, Italy}
\email{fabio.lepiane@ismn.cnr.it}
\and
Matteo Baldoni
\institute{Istituto per lo Studio dei Materiali Nanostrutturati (ISMN)}
\institute{Consiglio Nazionale delle Ricerche (CNR)\\
Bologna, Italy}
\email{matteo.baldoni@ismn.cnr.it}
\and
Mauro Gaspari
\institute{Alma Mater Studiorum}
\institute{Universit\`{a} di Bologna\\
Bologna, Italy}
\email{mauro.gaspari@unibo.it}
\and
Francesco Mercuri*
\institute{Istituto per lo Studio dei Materiali Nanostrutturati (ISMN)}
\institute{Consiglio Nazionale delle Ricerche (CNR)\\
Bologna, Italy}
\email{francesco.mercuri@cnr.it}
}
\def\titlerunning{MAMBO: Materials And Molecules Basic Ontology}
\def\authorrunning{F. Le Piane, M. Baldoni, M. Gaspari, F. Mercuri}
\begin{document}
\maketitle
\begin{abstract}
Recent advances in computational and experimental technologies applied to the design and development of novel materials have brought out the need for systematic, rational and efficient methods for the organization of knowledge in the field. 
In this work, we present the initial steps carried out in the development of MAMBO - an ontology focused on the organization of concepts and knowledge in the field of materials based on molecules and targeted to applications. 
Our approach is guided by the needs of the communities involved in the development of novel molecular materials with functional properties at the nanoscale. 
As such, MAMBO aims at bridging the gaps of ongoing efforts in the development of ontologies in the materials science domain. 
By extending current work in the field, the modular nature of MAMBO also allows straightforward extension of concepts and relations to neighboring domains. 
Our work is expected to enable the systematic integration of computational and experimental data in specific domains of interest 
(nanomaterials, molecular materials, organic an polymeric materials,
supramolecular and bio-organic systems, etc.). 
Moreover, MAMBO can be applied to the development of data-driven integrated predictive frameworks for the design of novel materials with tailored functional properties.
\end{abstract}

\section{Introduction}
The progress of several relevant fields in science and technology is often related to the design and development of novel materials. 
Accordingly, advancements in the materials development domain are considered key enablers for innovation in application fields of great technological and socio-economic relevance\cite{KeyPolicy}.
Recent developments of data-centric technologies have empowered significant progress in a very broad range of application sectors\cite{2021DataApplications,QIN2012220}.
As for many other fields, these advancements have had a significant impact also on research and innovation for materials\cite{Himanen2019Data-DrivenPerspectives,Li2019ADomain,Pollice2021}.

This trend has also been boosted by the outstanding breakthroughs in multiscale materials modelling and in applied data-intensive approaches\cite{Agrawal2016Perspective:Science,LePiane2020PredictingLearning}. In particular, advances in high-performance and high-throughput computing (HPC/HTC) and data-driven technologies, including artificial intelligence, have further accelerated the process.\\
The approaches currently pursued by the community involved in the process of design and development of materials leverage the integration of both computational and experimental methods. Computational techniques in the materials development domain cover a broad range of methods and approaches, from electronic structure calculations to continuum (full-scale) simulations\cite{Rosso2017WhatVersion}. Multi-scale techniques are generally applied to bridge the knowledge about materials across a wide range of spatial and temporal scales. From the experimental point of view, a huge variety of methodologies is commonly used to gather information about materials throughout the development process. These approaches lead typically to a very large amount of unstructured and uncorrelated information on materials.\\
The tremendous increase of the growth rate of data related to materials development has therefore led to the need for an organization and structuration in the field. Moreover, the application of FAIR (Findable, Accessible, Interoperable, Reusable) concepts to data that are relevant to the domain of materials development will trigger new paradigms\cite{Wilkinson2016Comment:Stewardship}. Recent work has demonstrated the relevance of FAIR principles for the automatic retrieval of information or extraction of knowledge from materials data. These efforts are expected to put forward cutting-edge and efficient technologies for using/reusing data on materials and to support the development of data-centric predictive models for innovation\cite{Gygli2020SimulationModeling,Draxl2018NOMAD:Science}. \\
Ontologies constitute a valuable and powerful tool to address the issues related to the organization of knowledge within a given domain. Although still in its early stages, the development and application of ontologies in the materials domain is already displaying the tremendous potential of this approach\cite{Ashino2010MaterialsKnowledge,Cheung2008TowardsMaterials}.
In consequence of this emerging interest, recent research and cooperation activities have addressed the development of ontologies targeted to materials.
Cooperative efforts in the development of ontologies and semantic technologies for materials are expected to enable the implementation of efficient platforms for the organization of materials data or the realization of complex workflows in the research and innovation process\cite{Horsch2020OntologiesMarketplace}.
These steps can be considered key enablers for the digitalisation of the materials development process at several levels. \\
A significant amount of work focused on the development of top- and middle-level ontologies for materials. The most relevant contribution to the field has been provided by the development of the European Materials Modelling Ontology (EMMO)\cite{Horsch2020ReliableOntology}, which is still in progress. Beside that, domain ontologies for specific use cases related to materials also began to be developed\cite{Li2020AnDomain}. Indeed, the broad scope of the research on materials requires an extensive work covering a manifold of different aspects and knowledge. In this context, recent work has addressed specific application domains, such as simulations of crystalline materials or single-molecule systems\cite{Li2020AnDomain, Li2019ADomain, Degtyarenko2008ChEBI:Interest}. 
However, attempts in the organization of knowledge focused on materials where aggregation properties at the molecular level are relevant seem to still be lacking.\\
In this work, we introduce MAMBO - the Materials and Molecules Basic Ontology. MAMBO is focused on concepts and relations emerging on materials where the relationship between individual molecules and molecular aggregation is relevant to the properties of the system. This is the case, for example, of molecular materials, nanomaterials, supramolecular materials, molecular thin-films and other similar systems of interest. These materials play a crucial role in several application fields and technologies, including organic electronics and optoelectronics (OLEDs, organic thin-film transistors), organic and hybrid photovoltaics (organic and perovskite solar cells), bioelectronics (neural and brain interfaces), molecular biomaterials, and several others.\\
In an essentially modular structure, MAMBO can easily be extended to cover other aspects of the domain considered (for example, introducing new computational methods for molecular materials) or to neighboring domains. Concepts and relations introduced in MAMBO could also possibly be used to develop new ontologies for modelling specific tasks, integrating them with top-level ontologies in order to provide full-level interoperability between different applications and computational methods. The development of MAMBO can lead to efficient tools for retrieving and analysing computational and experimental information in the development of materials based on molecular systems. Moreover, MAMBO can provide the basis for the implementation of data-driven technologies and workflows, for example based on machine learning, for the design and development of novel functional materials.

\section{Related Work}
Previous work has focused on the development of ontologies in the materials domain, focusing on different aspects of the problem and at different levels of details. As stated before, EMMO constitutes one of the most significant efforts in the field. EMMO aims at developing a general ontology for materials and modelling\cite{Ghedini2017EMMOONTOLOGY}. 
Other ontologies related to MAMBO address more specialized sub-topics in the domain of materials. ChEBI (Chemical Entities of Biological Interest) is an ontology focused mostly on chemical systems\cite{Degtyarenko2008ChEBI:Interest} .
Despite the very focused scope, several concepts introduced in ChEBI can be
related to the domain covered by MAMBO.
A very recent effort led to the Materials Design Ontology (MDO), which defines concepts and relations to cover knowledge in the field of materials design and especially in solid-state physics\cite{Li2020AnDomain}. 
Moreover, other ontologies developed in the framework of digitalisation and virtualisation can be related to MAMBO. These include OSMO (ontology for simulation, modelling, and optimization), and ontologies developed within the European project VIMMP (Virtual Materials Marketplace Project)\cite{Horsch2020OntologiesMarketplace,2020VIMMPMarketplace}. In addition, MAMBO also aims at connecting with other efforts focused on the development of materials databases (OPTIMADE, NOMAD)\cite{2021TheOPTIMADE,2021TheNOMAD,Draxl2018NOMAD:Science}.

\section{Typical application scenarios}
The application scenarios addressed by MAMBO focus on frameworks occurring in the development of the class of molecular materials and related systems. The main application scenarios we have selected, stemming from potential use cases, are:
\begin{itemize}
    \item The retrieval of structured information from databases on molecular materials.
    \item The definition of complex workflows for the modelling of systems based on molecular materials.
\end{itemize}
These scenarios were defined on the basis of the analysis of current work in the field of ontology development in the materials science domain and discussions with domain experts.
For example, MAMBO will enable semantic searches in databases on multi-scale modelling and characterization data on OLEDs\cite{Andrienko2020,Baldoni2018SpatialDynamics}. Here, data can include information about the basic chemico-physical entities constituting parts of the active systems (for example molecules, polymers, etc.), aggregated systems and full-scale devices. In other applications, MAMBO can be used to model the steps of a complex computational workflow which addresses a specific scientific/technological problem in the framework of molecular materials. For example, the modelled workflows can lead to the implementation of efficient computational approaches for the screening of properties of a given class of molecular systems. The data obtained by simulations can further be used to implement predictive data-driven models, for example for designing novel materials. The semantic interoperability provided by MAMBO will enable the integration between simulation data, possible integration with data stemming from other sources (experiments, characterization, etc.) and the application of data-centric approaches.

\section{Basic principles, development process and methods}
MAMBO was designed and developed to address the challenges in material science we have highlighted. 
We started with in-depth discussions and meetings between the MAMBO development team and domain experts to define a set of possible application scenarios and use cases. These steps led us to the definition of:
\begin{itemize}
    \item A set of competency questions, that is a set  of typical questions for which the information in MAMBO should provide answers.
    \item A set of typical tasks, which will be supported by the MAMBO ontology.
    \item A set of typical use cases, as in the examples outlined above.
\end{itemize}
Due to the peculiar nature of the typical development approaches pursued in the considered application area, we modelled the main concepts of the ontology associating them to specific problem solving methods (PSMs)\cite{Fensel2003TheUPML}. 
PSMs can be used to define operations to be performed to accomplish a given goal, related to a task. According to standard PSMs development approaches, complex tasks, defined by specific use-cases, were decomposed into subtasks. In each task and subtask, the required set of pre- and post-conditions was also defined.  This approach helped us to individuate relevant terms and connections between concepts in the considered application scenarios and use cases.
A wide set of different techniques was used, aimed at catching relevant terms and concepts on the basis of the textual content of the discussions.  In this way, we have individuated an initial set of terms to be used as a ground basis for MAMBO.\\
We interviewed experts in many different sub-domains of the
general fields related to the main MAMBO topics (researchers and
professionals working in the field of molecular and
nanostructured materials and their applications).
The collection of information involved asking the experts
a general description of their research work, also identifying the most crucial terms and concepts without which they are unable to describe or define their activities.
This step allowed us to identify the main common concepts 
used and to annotate an initial list of relevant terms. 
An initial group of 5 experts was initially involved with
in-depth interview taking place over the course of several
days.
This work is still in progress, involving a much larger
group of experts. Namely, about 80 experts have been
contacted for a survey to be used in the next development stage.
From the terms obtained in these first steps, an initial representation of the concepts and relations was drafted. In more detail, a “hybrid” approach (bottom-up and top-down) was used, to better represent the different nature of concepts involved in the development of the MAMBO ontology. A tentative set of relationships among terms was initially built and adjusted iteratively. Further details about the development process of MAMBO will be provided in a future work.

AMB\subsection{Integration with other ontologies}
Beside the interaction with domain experts and knowledge engineers, one of the steps considered in the development of MAMBO concerns the possible integration with other ontologies. The integration will also involve reuse of concepts and terms from other ontologies. In particular, the specific domain of application of MAMBO suggests connections with the following ontologies:
\begin{itemize}
    \item EMMO - a reference ontology with possibile links at the upper level
    \item ChEBI - connections with MAMBO on concepts related to individual molecules
    \item MDO - integration between crystalline (typically inorganic) systems and molecular (organic) materials.
\end{itemize}
To approach this problem, we started a conceptualization process based on the efforts made in the development of these ontologies and, when needed, we partly redefined
some of the concepts on the basis of the specific target domain.
We stress that, at the current stage of development, 
we are focusing on the conceptualization of entities and
relationships in the domain of interest,
trying to capture relevant concepts in a very complex field.
The full formalization of MAMBO will follow this step, and
will be the subject of a forthcoming paper.
Acoordingly, the formalization of the integration with other ontologies and of the reuse in terms of concepts and relationships in MAMBO is still in progress.

\section{Realization of MAMBO}
To address the design principles outlined above, we started the design of MAMBO with an essentially modular structure in mind. According to this set-up, we defined  a set of core concepts and basic relationships, which are connected to other lower-level hierarchies, as it will be shown below.

\subsection{Main concepts defined in MAMBO}
\begin{figure}[!ht]
	\centering
	\includegraphics[scale=0.3]{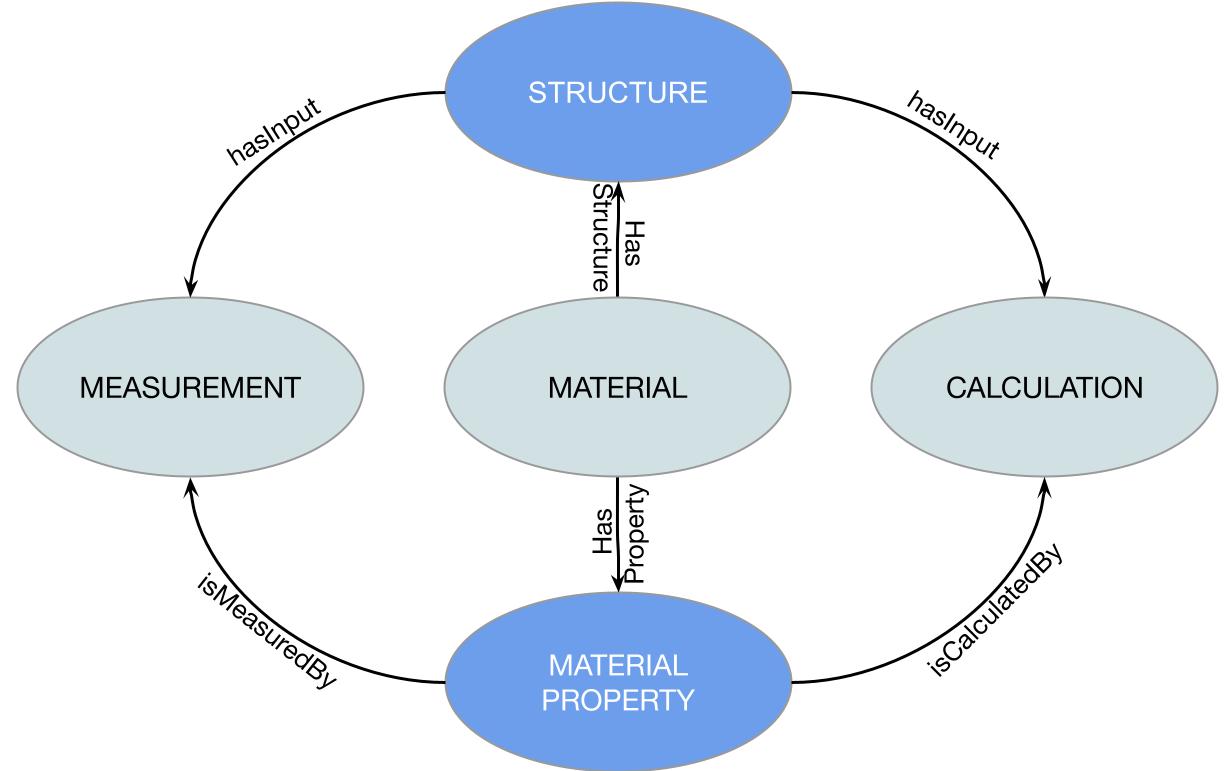}
	\caption{Core concepts of MAMBO: the ontology revolves around the concepts of {\fontfamily{tt}\selectfont Material}, {\fontfamily{tt}\selectfont Calculation} and {\fontfamily{tt}\selectfont Measurement}. An object ({\fontfamily{tt}\selectfont Material}) is represented by its structural features and properties, while {\fontfamily{tt}\selectfont Computational} and experimental ({\fontfamily{tt}\selectfont Measurement}) workflows are connected through a common interface to {\tt Material Property}.}
	\label{core}
\end{figure}

The basic initial structure of MAMBO provides a generalisation of terms and relationships emerged throughout the development process. The main concepts will be represented by classes in the formalized ontology. In a nutshell, the output of this process  is the identification of the MAMBO “entities” (more specifically, materials entities), which are the objects of our investigations and can have a structure and/or a property. {\fontfamily{tt}\selectfont Structure} and {\fontfamily{tt}\selectfont Material Property}\footnote{It must be noted that {\tt Material Property} should not be intended as a concept taken from particular ontology languages but it is a proprietary definition of MAMBO, opt to represent the concept of "property" in the chemical/materials science realm.} can therefore be defined as concepts related to the main class {\fontfamily{tt}\selectfont Material}. Other relevant concepts are {\fontfamily{tt}\selectfont Calculation} and {\fontfamily{tt}\selectfont Measurement}. The scheme of the main concepts defined in MAMBO is shown in Fig. \ref{core}. 

\subsection{Drafting the "Structure" class}
\begin{figure}[!ht]
	\centering

	\includegraphics[scale=0.24]{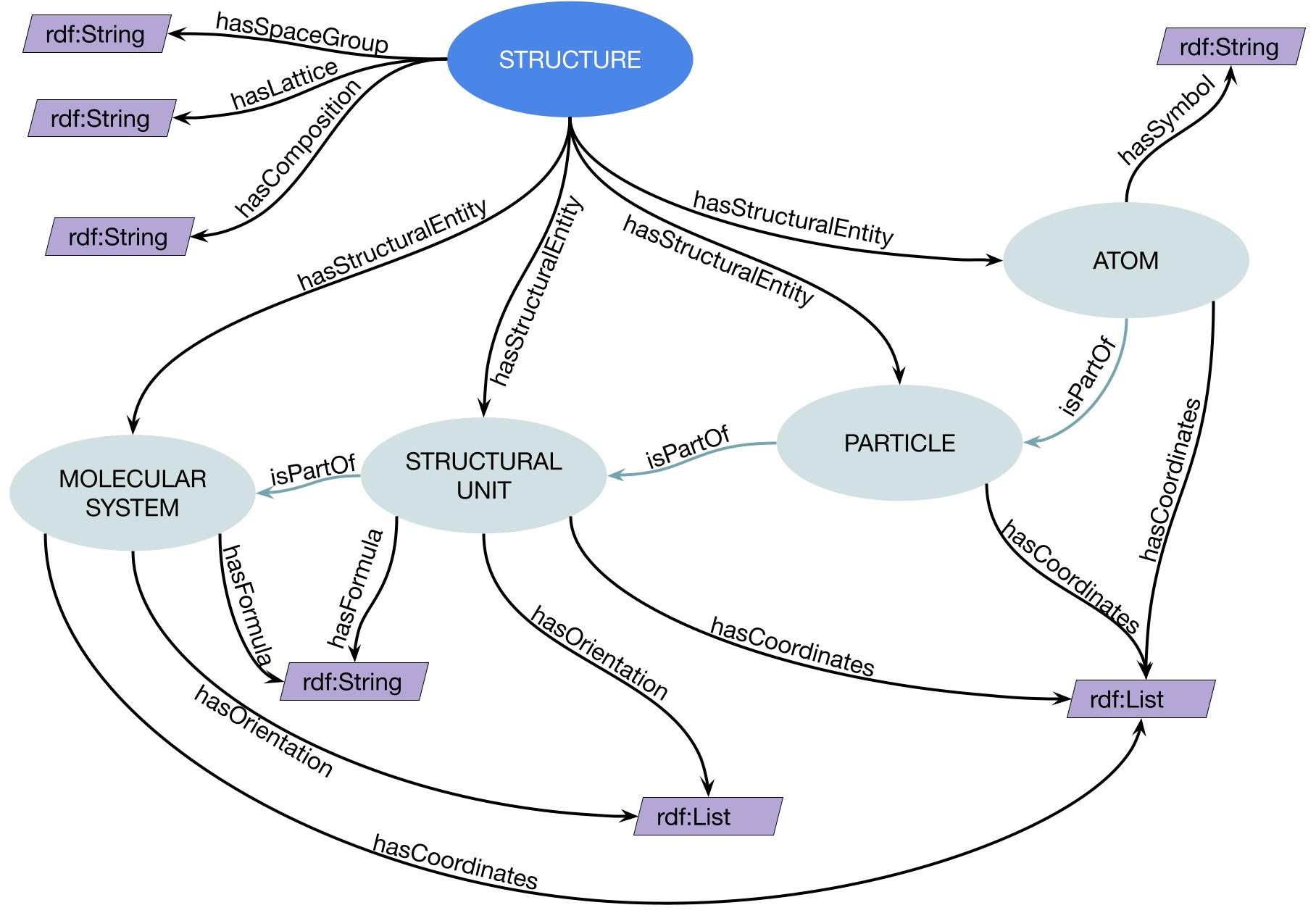}
	\caption{
	Draft scheme of the {\fontfamily{tt}\selectfont Structure} class. The main concepts and relationships used in the {\fontfamily{tt}\selectfont Structure} class are related to the analysis of actual workflows emerging from typical problem solving tasks involving molecular materials. Terms and relationships are connected to both computational and experimental techniques and methods.}
	\label{structure}
\end{figure}

As an example of the strategy pursued in the development of MAMBO, we briefly discuss the initial draft of the {\tt Structure} class. Some of the concepts and relationships identified are shown in Fig. \ref{structure}.
This schema aims at providing a semantic asset for the organisation of knowledge concerning the structure of systems based on molecular materials. Generally speaking, the {\fontfamily{tt}\selectfont Structure} class relates to the structural property (in 3D space and time) of an object ({\fontfamily{tt}\selectfont Material}) that can be measured and/or simulated. In the context of molecular materials, we found it useful to consider a structure as made by “structural entities”, which can have different features. A structural entity can be a molecular system (for example, a molecule), a molecular structural unit (for example, a functional group/subgroup), a particle or an atom. On the basis of these definitions and of the analysis performed in the previous steps, we started to define lower-level concepts and attributes. In particular, our effort aims at generalising and extending some of the cases considered in other similar ontologies. 
For example, the position in space of a molecular system can be defined through the concept of “orientation”, which can have a rotation matrix or a quaternion vector as properties defining the actual molecular orientation. 
The {\tt Structure} class can have properties that are related to the object (simulated or measured) as a whole, for example defining the periodicity of the system.
Clearly, in Fig. 2 only a subset of the full set of required concepts and relationships is shown.
Initial instantiation tests on different use cases provided encouraging results. For example, let us consider the case of a simulation workflow involving a liposome in a water solution. One of the main objects of our investigation will be the liposome structure, which can be considered as a particular shape of a lipid bilayer. Therefore, the liposome will be the instance of the {\tt Structure} class. The particular phospholipid constituting the liposome bilayer (for example, 
dipalmitoylphosphatidylcholine or DPPC) will be an instance of the {\tt Molecular System} class. 
A phosphate group is a possible instance of {\tt Structural Unit}. Information about this class can for example be useful in some simulation methods (molecular dynamics, etc.). A phosphorus atom can be an instance of the {\tt Atom} class. In the same example, the water solution surrounding the liposome (and contained within the liposome cavity) can be considered as another instance of the {\tt Structure} class.
The analysis of use cases is currently in progress to extend the scope of the {\tt Structure} class and of other classes shown in Fig. \ref{core}.

\subsection{Drafting the "Property", "Measurement" and "Calculation" classes}
\begin{figure}[!ht]
	\centering
	\includegraphics[scale=0.5]{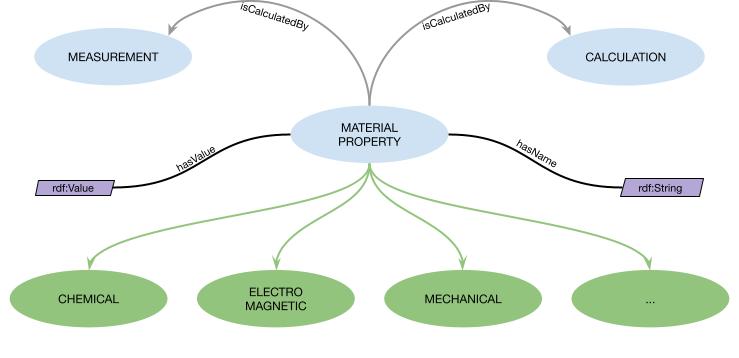}
	\caption{
	Scheme of the {\tt Material Property} class. This class is mainly described by name and value attributes.
	The {\tt Material Property} class is also symmetrically linked to {\tt Measurement} and {\tt Calculation} classes. 
	}  
	\label{property}
\end{figure}

\begin{figure}[!ht]
	\centering
	\includegraphics[scale=0.5]{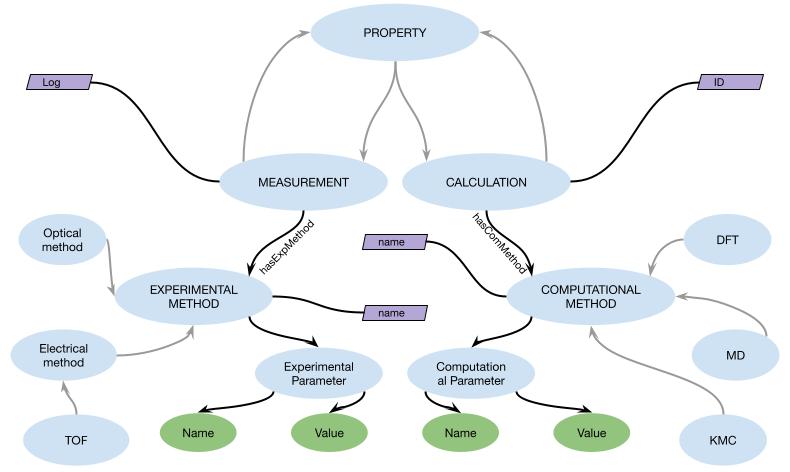}
	\caption{
	Scheme of the {\tt Measurement} and {\tt Calculation} classes. They both have their respective "method" class ({\tt Experimental Method} and {\tt Computational Method}, respectively) which lead to the different experimental and computational methods, while gathering their parameters.
	}
	\label{meas_calc}
\end{figure}

We then proceeded defining the bigger picture, focusing in particular on three classes, {\tt Property}, {\tt Measurement}
and {\tt Calculation}, investigating their mutual relationships.
These three classes are deeply linked: the latter two act as a middle ground that let us treat experimental and computational results in all their specificity, merging then the results into {\tt Property}. 
This allow us to convey the general characteristics (i.e. both computationally and experimentally accessible)
of a system into the more general {\tt Property} class, while keeping more specific attributes in the
{\tt Measurement} and {\tt Calculation} classes.
In turn, {\tt Measurement} and {\tt Calculation} classes
are connected to a specific class for their corresponding method ({\tt Experimental Method} and {\tt Computational Method}, respectively) which represent the set of
different methodologies and their respective parameters.\\
These relationships are represented in Fig. \ref{property} and Fig. \ref{meas_calc}.

\subsection{Formalization and implementation procedures}
Current steps in the development of MAMBO concern the initial formalisation of relationships between concepts.
This strategy is initially applied to the classes shown in Fig.
\ref{core}, and progressively extended to include a more structured representation.
To this end, we used the OWL 2 language\cite{W3COWLWorkingGroup2012OWLOverview}, 
while we are evaluating the possibility of re-implementing MAMBO with the OWLReady framework and library\cite{Lamy2017Owlready:Ontologies}.\\
At the moment, we are using the RDF/XML syntax.
First of all, we drew the informal representation of a module of the ontology, trying to define the relationships between different concepts and trying to identify the main properties of each class and subclass.
This step also involved extracting the main hierarchical relationships between classes, identified according to the hybrid (top-down/bottom-up) approach mentioned above.
Once we reached a results that we felt to be consistent, we then shifted to the formal implementation in OWL to see if the proposed schema can be effectively modelled in a formal way. 
In this initial implementation stage, we applied OWL constructs to define the relationships between classes.
Moreover, instantiation tests allowed us to validate the relationships between the defined classes.
As expected, our reasoning turned out to be slightly imperfect but the conceptual structure of the actual implementation was almost identical to that of the informal scheme. In particular, we used the \textit{WebProt\'{e}g\'{e}} editor\cite{Tudorache2013WebProtege:Web}
for editing the project interactively and collaboratively.\\
At the time of this writing, we implemented a testing version of the {\fontfamily{tt}\selectfont Core} and {\fontfamily{tt}\selectfont Structure} hierarchies shown in Fig. \ref{core} and Fig. \ref{structure}, respectively. 
However, we stress that the formal implementation
of MAMBO is still at its early stages, and
the relationships and classes shown in the
aforementioned Figures are susceptible to changes.
Also, we are currently setting up a consistent naming standard, able to
support and express the semantic relationships involved in the
ontology modelling.
An initial version of the MAMBO ontology is
available on GitHub\footnote{https://github.com/egolep/MAMBO}.

\section{Future steps}
The development of MAMBO is still in progress and will require several additional steps.
As mentioned above, we are currently working at the formalization and implementation of the basic structure of the ontology, encoding the main concepts and hierarchies. This initial implementation step will also guide us in defining a set of useful relationships between classes. These relationships will possibly be reused for other hierarchies in MAMBO. This work will also allow us to assess the possibilities offered by different implementation strategies.
Another development step will concern the extension of MAMBO to cover specialized domain areas. This extension process will be focused on the classes shown in Fig. 1. One of the ambitions of MAMBO is to organize in a comprehensive formal way the computational and experimental knowledge emerging from research on molecular materials. As such, MAMBO will target a broad range of concepts and relationships in the domain, ranging from multiscale computational methods to experimental characterization of the specific class of materials considered. 
Tho address this general goal, however, we plan to reuse concepts and terms from other ontologies, defining new relationships targeted to the specific use cases and tasks in a functional way.
Finally, we will carry out a thorough evaluation of MAMBO, targeted especially to the assessment of the performance in terms of tasks defined by specific use-cases. 

\section{Conclusions}
In summary, we introduced MAMBO - an ontology focused on the representation of the terminological knowledge relevant for applications and computational processes involving materials based on molecules and similar systems.
The knowledge modelled in the proposed ontology relates to both computational and experimental information about molecular materials and related systems, providing a fully interoperable platform. The ambition of MAMBO is also to model a broad range of concepts and relationships of common use in the field. These include for example methods and approaches involved in the multiscale modelling of molecular materials. Treating empirical and computational information on materials on the same footing will also enable a full integration of data. Beside the realization of a platform for the organization of pre-evaluated data, MAMBO will also find application in the definition of computational, experimental or integrated workflows targeted to specific tasks. The approach pursued in the development of MAMBO will allow the extension of the semantic asset towards other similar fields of interest in the domain of molecular materials. Moreover, the concepts and relationships defined within MAMBO can also be structured in the framework of other top-level ontologies. Although still in an early development stage, our initial assessment and instantiation tests demonstrate the potential of MAMBO in the specific field of molecular materials and nanostructures.  Work is in progress to implement the formal structure of MAMBO, to extend the scope of classes and to test performance and use in applications.

\printbibliography
\end{document}